# Experimental probe of band structures of bilayer valley photonic crystals

Xiang-Fei Guo[†], Jian-Wei Liu[†], Hong-Xiang Chen, Fu-Long Shi, Xiao-Dong Chen[*], Jian-Wen Dong

*School of Physics & State Key Laboratory of Optoelectronic Materials and Technologies,*
*Sun Yat-sen University, Guangzhou 510275, China.*

[†]These authors contributed equally to this work
[*]Corresponding author: chenxd67@mail.sysu.edu.cn

**Research on two-dimensional van der Waals materials has demonstrated that the layer degree of freedom can significantly alter the physical properties of materials due to the substantial modification of bulk bands. Inspired by this concept, layered photonic systems have been proposed and realized, revealing novel phenomena absent in their monolayer counterparts. In this work, we experimentally investigate the band structures of bilayer valley photonic crystals. Two typical structures with different stacking configurations are experimentally imaged via the near-field scanning technology, exhibiting distinct bulk band structures. Furthermore, different topological edge modes induced by distinct topology are observed, revealing that the layer degree of freedom can be regarded as a pseudospin and offer further capabilities for controlling the flow of light. Our work not only elucidates the evolution of band structures from monolayer to bilayer topological systems but also provides an experimental platform for the further exploration of bilayer topological insulators.**

## I. INTRODUCTION

Over the past few decades, stacking or twisting bilayer van der Waals materials have garnered significant attention due to their remarkable electronic properties [1-9]. For instance, the semiconductor $MoS_2$ transitions from a direct to an indirect band gap material when shifting from a monolayer to a bilayer [10,11]. Similarly, twisted bilayer graphene exhibits zero Fermi velocity (i.e. flat band) at the so-called magic angle, leading to superconductivity through the formation of moiré superlattices [12,13]. Compared to their monolayer counterparts, the addition of a second layer in bilayer van der Waals materials significantly modifies the bulk band structures, markedly altering their physical characteristics. This concept has recently been applied to photonic systems, leading to the development of one-dimensional (1D) and two-dimensional (2D) photonic bilayer lattices, which enable the realization of asymmetric radiation effects [14,15], slow-light phenomena [16,17], light localization [18-20], and tunable resonant chiral behaviors [21]. Among all the physical characteristics, topology, which considers the global property of bulk band structure, has attracted much attention [22-30]. One prominent example is the topological photonics, which holds promise for various applications due to topologically protected edge modes that enable reflection-free light propagation [31-46]. In a topological system, the topological phase transition is typically accompanied by the change of bulk band structure, i.e. the closing and reopening of band gaps. Additionally, studies on the open systems have revealed that the bulk band spectra topology can enable richer topological properties, such as the non-Hermitian skin effect [47-52]. Therefore, probing the bulk band structure is a significant in the study of topological systems.

In addition to the extensive study of 2D topological phases in monolayer systems, new topological effects induced by the layer degree of freedom have been explored in bilayer photonic and sonic crystals. By treating the layer degree of freedom as a pseudospin, intriguing phenomena such as gap width tunning and spin-layer locking effects have been observed [53-55]. The interlayer coupling in bilayer system strongly influences the bulk band structure, leading to phenomena distinct from those in monolayer systems. For instance, in bilayer photonic crystals, the coupling between layers can create new pathways for light propagation, enhancing or suppressing certain modes depending on the alignment and spacing of the layers. This can be exploited to engineer photonic band gaps with greater precision, enabling more efficient control over light-matter interactions. Similarly, in bilayer sonic crystals, the interplay between layers can lead to the emergence of new acoustic modes and the modulation of sound wave propagation, offering potential applications in acoustic insulation and waveguiding. Despite these promising theoretical advancements, experimental studies on bilayer topological photonic systems remain limited. One of the primary challenges is the complexity of fabricating and characterizing bilayer structures with the necessary precision to observe these effects. Additionally, probing the bulk band structures of bilayer systems requires sophisticated experimental setups that can accurately image the excited field distributions.

In this work, we experimentally measure the band structure of a bilayer topological photonic system consists of two AA-stacking valley photonic crystal (VPC) slabs. Two typical bilayer VPCs with distinct topology are discussed. Both transverse-electric-like (TE-like) and transverse-magnetic-like (TM-like) bulk bands are simultaneously probed by means of microwave near-field scanning technology. Our experimental results demonstrate that the introduction of the second layer offers greater dispersion engineering capabilities compared to monolayer counterparts. Particularly, bulk bands of inversion-symmetric bilayer VPCs exhibit frequency-degenerate nodal points at the K point, a fragile topological feature protected by the non-trivial Euler class [56,57], are confirmed in our experiment. Furthermore, the layer-mixed and layer-polarized edge modes are observed in the experiment, confirming that the layer can be considered as an additional pseudospin in this bilayer system. Our work not only experimentally reveals the evolution of the band structure from monolayer to bilayer topological systems, but also provides an experimental platform for further exploration of underlying physics about non-Abelian and fragile phases in bilayer systems.

## II. BULK BAND EVOLUTION: FROM MONOLAYER TO BILAYER

### A. Structures and Hamiltonian analysis

We examine a bilayer topological structure comprising two AA-stacking VPC slabs, separated by a distance of $h_s$ [Fig. 1(a)]. Initially, if two slabs are placed far apart, they function independently without interlayer coupling, each supporting its own photonic band structure. However, as two VPCs are brought closer, the modes in each VPC slab start to couple, leading to a modification of the band structure due to the nonzero interlayer interaction. Notably, this band modification is highly sensitive to the relative morphology of the slabs, providing capabilities for dispersion engineering [see Fig. 1(c)]. To better understand the differences between band structures of monolayer and bilayer systems, we first analyze the bulk band structure of the monolayer VPC, and then consider two representative bilayer VPCs with distinct symmetry.

We consider a monolayer VPC whose inversion symmetry is broken by incorporating two triangular air holes with different sizes [middle panel of Fig. 1(b)]. This modification opens the gapless Dirac cone around the K or K' point within the Brillouin zone, resulting in a band gap [middle panel of Fig. 1(c)]. The gapped Dirac cone is well described by an effective low energy Hamiltonian:

$$H(k_x, k_y) = vk_x\sigma_x\tau_z + vk_y\sigma_y\tau_0 + m\sigma_z\tau_0, \tag{1}$$

where $v$ is the Fermi velocity, $m$ is the mass term representing the staggered sublattice potential at two inequivalent K and K' valleys. The Pauli matrices $\sigma_i$ and $\tau_i$ ($i = x, y, z$) function in the subspace of sublattice and valley, respectively. The third term in Eq. (1) is responsible for the gapped bulk band structures, which align with the theoretically predicted gapped Dirac cones.

When the bilayer VPC is constructed by two monolayer VPCs, the bilayer system can exhibit either mirror symmetry or inversion symmetry depending on the stacking configuration. In the case of a mirror-symmetric (MS) bilayer VPC [left panel of Fig. 1(b)], the effective low energy Hamiltonian around K or K' points can be expressed as

$$H(k_x, k_y) = vk_x\sigma_x\tau_z s_0 + vk_y\sigma_y\tau_0 s_0 + m\sigma_z\tau_0 s_0 + t_\perp\sigma_0\tau_0 s_x, \tag{2}$$

where $t_\perp$ is the interlayer coupling and the Pauli matrices $s_i$ function in the layer subspace. Since the signs of the mass term are identical in two layers [left inset in Fig. 1(c)], all bulk bands in this case are separated from each other [left panel of Fig. 1(c)], similar to the monolayer configuration. These four separated bulk bands can also be inferred from an analysis of the bilayer unit cell's symmetry. All symmetry operations that keep the mirror-symmetric bilayer unit cell invariant constitute the point group $\mathcal{M}_1 = D_{3h}$. The subgroup of $\mathcal{M}_1$ that leaves the wave vector K = ($4\pi/3a$, 0) invariant is $\mathcal{M}_K = C_{3h}$. Since the $C_{3h}$ point group does not support degenerate modes, all bulk modes at the K point exhibit different frequencies, resulting in four separated bulk bands.

In contrast, in an inversion-symmetric (IS) bilayer VPC [right panel of Fig. 1(b)], the effective Hamiltonian can be expressed as

$$H(k_x, k_y) = vk_x\sigma_x\tau_z s_0 + vk_y\sigma_y\tau_0 s_0 + m\sigma_z\tau_0 s_z + t_\perp\sigma_0\tau_0 s_x. \tag{3}$$

Now, the signs of the mass term are opposite in two layers [right inset in Fig. 1(c)]. Different from the previous bilayer case, two nodal points emerge between the upper two bands and the lower two bands, respectively [right panel of Fig. 1(c)]. The degeneration of bulk modes at the K point can also be inferred from the analysis of the bilayer unit cell's symmetry. All symmetry operations that keep the inversion-symmetric bilayer unit cell invariant constitute the point group $\mathcal{M}_2 = D_{3d}$. The subgroup of $\mathcal{M}_2$ that leaves the wave vector K invariant is $\mathcal{M}_\mathbf{K} = C_{3v}$. Because the $C_{3v}$ point group support degenerate modes, we expect symmetry-protected degenerate bulk modes at the K point.

**B. Monolayer VPC**

As illustrated in Fig. 2(a), the fabricated monolayer VPC is a dielectric slab with relative permittivity $\varepsilon_r = 8$ and a thickness of 5 mm. This slab exhibits a honeycomb array of equilateral triangular air holes. Each unit cell includes two air holes, with lateral dimensions $s_A = 3.6$ mm and $s_B = 8.4$ mm. Note that this VPC slab is primarily designed for TE-like modes, but we include a discussion of TM-like modes for completeness and accuracy. As TE-like and TM-like modes coexist within the VPC slab, both TE-like and TM-like bands are consequently observed in the subsequent simulations and experiments. To investigate the bulk band structure of the monolayer VPC, we employ a source antenna placed beneath the sample to excite the electromagnetic field. We then use another antenna to scan the electric field on the sample's surface [Fig. 2(b)]. These antennas are intentionally bent to enhance the detection of in-plane fields. Additionally, we scan the electric field profiles on the upper surface of the VPC slab, allowing us to simultaneously capture both the TE-like and TM-like modes in our experiment. This approach marks a novel addition to similar studies, as previous experimental results typically focused on measuring either TE-like or TM-like band dispersions individually. Additional details regarding the experimental setup can be found in the Supplementary Figs. S1 and S2. The upper row of Fig. 2(c-e) displays the measured $E_x$ fields at three different frequencies. After acquiring these field profiles, a Fourier transformation is applied to obtain the equifrequency contours in momentum space for each frequency [lower rows of Fig. 2(c-e)]. The simulated contours of the TE-like and TM-like modes are outlined by green and blue dashed lines, respectively. It is evident that both measured TE-like and TM-like modes closely match the simulation results. In the experiment, both the source and probe antennas are bent, leading them to respond differently to TE-like bulk modes propagating along the $x$ and $y$ directions. This results in the observed anisotropy of the measured bulk bands. By putting the equifrequency contours along the frequency dimension, the whole 2D bulk band structure is constructed. The bulk band structure along high-symmetry lines in the first Brillouin zone is shown in Fig. 2(f), further confirming our ability to simultaneously and accurately probe both TE-like and TM-like band structures. In the bulk band, we observe a band gap for the TE-like modes around 9 GHz. Notably, the band extrema are located at the K and K' valleys, which is a key feature of VPCs.

**C. Bilayer VPC**

Based on the bulk band structure of the monolayer VPC, we now investigate the bulk band structures of the bilayer VPC. To simplify the system without losing generality, we ensure that both the upper layer (Layer 1) and the lower layer (Layer 2) share the same structural parameters. Both layers have the same lattice constant ($a_1 = a_2 = 12$ mm), the same thickness ($h_1 = h_2 = 5$ mm), and the same lateral dimensions for the air holes. To support the entire system, a foam spacer membrane ($\varepsilon_r = 1.06$) is placed between the two VPC layers in the experiment.

As previously discussed, the bulk band structure of the bilayer system depends on the relative morphology of two slabs. We first consider a bilayer configuration with mirror symmetry, where the lateral dimensions of the air holes are $s_{A1} = s_{A2} = 3.6$ mm, $s_{B1} = s_{B2} = 8.4$ mm [Fig. 3(a)]. Similar to the experimental setup for the monolayer measurement, we place a bent antenna beneath Layer 2 to excite the electromagnetic field and use another bent antenna to scan the electric field profile above the upper surface of Layer 1. By applying Fourier transformation to the measured field profiles, we obtain the bulk band structure of this MS bilayer VPC, as shown in Fig. 3(b). The bulk bands of the TE-like modes are gapped between the second and third bands. Compared to that of the monolayer system, this band dispersion can be seen as a superposition of two sets of monolayer bands. To further validate this hypothesis, we conduct a numerical simulation of the MS bilayer VPC. Figure 3(c) displays the simulated $|H_z|$ fields and power fluxes at the central plane of Layer 1 and Layer 2 for four bulk modes at the K point. Considering mode ① and mode ② (mode ③ and mode ④), their $|H_z|$ fields and the power fluxes in both layers are identical. Mode ① and mode ② have their $|H_z|$ fields localizing at the air holes with large lateral size, while mode ③ and mode ④ at the air holes with small lateral size. In addition, mode ① and mode ② have opposite power fluxes to mode ③ and mode ④. As the power flux can be considered as the photonic orbit, mode ① and mode ② possess left-handed circular polarized (LCP) orbital angular momentum, while mode ③ and mode ④ possess right-handed circular polarized (RCP) orbital angular momentum. These eigenmode profiles closely resemble those in the monolayer VPC [see Supplementary Fig. S3], indicating that the MS bilayer VPC can be effectively viewed as a superposition of two monolayer VPCs.

In addition to the MS bilayer VPC, we investigate another bilayer configuration with inversion symmetry. In the IS bilayer VPC, the lateral dimensions of the air holes are $s_{A1} = s_{B2} = 3.6$ mm, $s_{B1} = s_{A2} = 8.4$ mm [Fig. 4(a)]. Similar procedures are followed to obtain the bulk band dispersion for IS bilayer VPC, as shown in Fig. 4(b). Similarly, there is a band gap between the second and third bulk bands of TE-like modes. However, unlike the MS bilayer VPC, the IS bilayer VPC features two nodal points in its bulk bands. One nodal point forms between the first two TE-like bands at the K point, while the other forms between the third and fourth TE-like bands. Moreover, as illustrated in Fig. 4(c), the mode profiles of two degenerate bulk modes at the nodal point differ significantly. For example, in the first two bands, two bulk modes have fields localized in different layers (Layer 2 for mode ① while Layer 1 for mode ②) and opposite power flux directions (clockwise for mode ① while counter-clockwise for mode ②). When the layer degree of freedom is viewed as a pseudospin and the power flux as the photonic orbit, these mode profiles show the feature of pseudospin orbital coupling. It also indicates that the modes in the IS bilayer VPC are not simply a superposition of the monolayer VPC modes.

### III. LAYER-MIXED AND LAYER-POLARIZED EDGE MODES

According to the bulk-edge correspondence, deterministic edge modes can be found at the domain wall between two topologically distinct insulators. For VPCs, which are the photonic analogues of quantum valley-Hall insulators, we typically use valley Chern numbers to characterize their topological properties. The valley Chern number can be calculated by integrating the Berry curvature within half of the first Brillouin zone around the K or K' point. Here, we focus on the valley Chern numbers at the K valley, while those at the K' valley can be easily obtained by considering the time-reversal symmetry.

First, we consider the domain wall constructed by two MS bilayer VPCs. A magnified view of the domain wall in Fig. 5(a) shows that Layer 1 and Layer 2 have the same boundary configuration. In the experiment, we place the source antenna beneath the Layer 2 domain wall and then probe the field profiles along the Layer 1 domain wall. The measured $|E_y|$ field at the frequency of 8.9 GHz is shown in Fig. 5(b). By further applying Fourier transformation, we can obtain the edge dispersion [Fig. 5(c)]. Since we place the source antenna at the left side of the domain wall, only the right-propagating modes with positive group velocity are well captured. As demonstrated in Fig. 5(c), there are two edge modes supported by this domain wall, which can be verified by difference of the valley Chern number over the domain wall. As we mentioned above, the MS bilayer VPC can be effectively regarded as the superposition of two monolayer VPCs. Therefore, it can be understood that there should be two valley edge modes with the same group velocity in the MS bilayer structure. These two valley edge modes can easily couple together, leading to a large-period oscillation of the electromagnetic field between the two layers [see Fig. 5(b)]. Therefore, we refer to these edge modes as the layer-mixed edge modes.

On the contrary, the IS bilayer VPC exhibits zero Berry curvature for every $k$-points in the Brillouin zone due to the combined inversion symmetry and time-reversal symmetry, resulting in a zero valley Chern number. However, recent studies reveal that the topological properties of such zero-Berry-curvature layered systems can be further characterized by the nonzero valley Euler numbers [56,57]. Notably, bulk bands at $k$-points far from the K or K' point deviate from those predicted by the effective Hamiltonian in Eq. (3). As a result, the valley Euler number is not strictly $\pm 1/2$. However, protected by the nonzero valley Euler numbers and the mirror symmetry $M_y: (x, y, z) \rightarrow (x, -y, z)$, the domain wall constructed by IS bilayer VPCs can still support topological edge modes. To see this, we construct the domain wall between two IS bilayer VPCs with opposite valley Euler numbers, showing different edge morphology at Layer 1 and Layer 2 [Fig. 5(d)]. The field profiles and the dispersion of the edge modes are measured in the experiment and shown in Fig. 5(e, f). Different from the layer-mixed edge modes, the filed profiles of these edge modes exhibit no oscillation with a large period. The edge modes are either localized in Layer 1 or in Layer 2. Since we only probe the field profiles on the upper surface of Layer 1, we only detect the right-propagating edge modes localized in Layer 1, as shown in Fig. 5(f). We refer to these edge modes as the layer-polarized edge modes. Regarding the layer as pseudospin, the layer-polarized edge modes can also be classified into pseudospin-up and pseudospin-down modes. For example, around the K valley at $k_x = -2\pi/3a$, the rightward edge modes are localized in the upper Layer 1, i.e., the rightward pseudospin-up edge mode. While the leftward modes are localized in the upper Layer 2, i.e., the leftward pseudospin-down edge mode. This dispersion configuration of the layer-polarized edge modes resembles the helical edge modes in quantum spin-Hall systems. Notably, topologically protected edge modes of MS bilayer VPCs or IS bilayer VPCs are robust against disorders or defects when intervalley scattering are suppressed. Therefore, defects such as inserting metal rods into the boundary will induce loss in the transmission, and see detailed discussion in Supplement D.

## IV. CONCLUSION

In conclusion, we have performed an experimental measurement of the band dispersion in an AA-stacking bilayer VPC. Two typical configurations, the mirror-symmetric bilayer VPC and the inversion-symmetric bilayer VPC, were discussed. By performing near-field scanning followed by Fourier transformation, we obtained the electromagnetic field distribution and the corresponding band structures. Our experimental results demonstrate that the additional layer can modify the bulk bands of monolayer systems. More significantly, under the nonzero interlayer coupling, this modification depends on the relative morphology of the two layers: the MS bilayer VPC can be effectively viewed as the superposition of two monolayer VPCs, while the IS bilayer VPC exhibits completely different bulk bands. The difference between their bulk band structure indicates that the MS bilayer VPC and the IS bilayer VPC have distinct topological properties, leading to different topological edge modes. Further, we observed the distinct layer-mixed and layer-polarized topological edge modes between the domain walls formed by MS and IS bilayer VPC, respectively. Our experimental results show that the layer degree of freedom offers significant potential for band engineering and novel light field modulation, which is absent in monolayer counterparts.

Our study not only illustrates the evolution of the band structure from monolayer to bilayer topological systems, but also provides an experimental platform for further exploration of bilayer topological insulators. Note that similar phenomena as demonstrated in our work could indeed be replicated in alternative structures, such as electric circuit boards. However, electric circuit boards often incorporate metal components that experience significant losses at near-infrared or visible frequencies. In contrast, our experimental platform is constructed entirely from dielectric materials, eliminating material losses at higher frequencies. This all-dielectric design also allows for scaling down the structure to nanometer dimensions, making it feasible to replicate the reported results within smaller scales. However, several potential challenges arise. Fabricating the experimental sample becomes challenging, as the triangular air holes will be subwavelength and about only hundreds of nanometers at terahertz frequencies, making precise fabrication critical. Furthermore, aligning two VPC slabs with nanometer accuracy presents additional fabrication difficulties. In terms of experimental measurement, employing near-field scanning techniques at near-infrared or visible frequencies can be complex, adding further obstacles to experimentation.

**Acknowledgements**

This work was supported by National Natural Science Foundation of China (12374364, 62035016, 12074443), Guangdong Basic and Applied Basic Research Foundation (2023B1515040023), Guangzhou Science, Technology and Innovation Commission (2024A04J6333), Fundamental Research Funds for the Central Universities of the Sun Yat-sen University (23lgbj021).

**Figures and Captions**

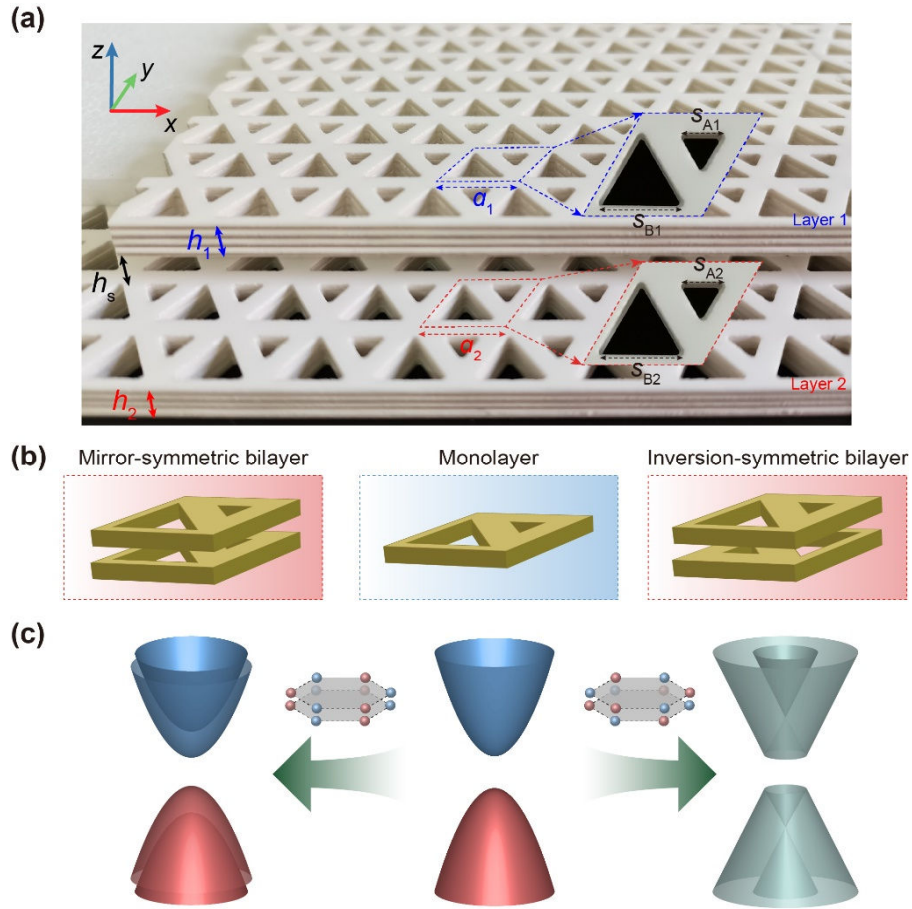

**FIG. 1. Bilayer VPC slabs and the corresponding bulk bands. (a)** Photograph of the bilayer VPC. Each VPC slab is a honeycomb lattice with the lattice constant $a$. Each unit cell includes two air holes, with lateral dimensions $s_A$ and $s_B$. Two slabs are separated by a space layer with the distance of $h_s$. **(b, c)** Schematics of unit cell and corresponding bulk band structures of monolayer VPC and bilayer VPCs with different stacking configurations. Notably, two bilayer VPCs have unit cells with distinct symmetry, resulting in completely different bulk band structures. Lattice structures of two AA-stacking honeycomb bilayers are shown as insets.

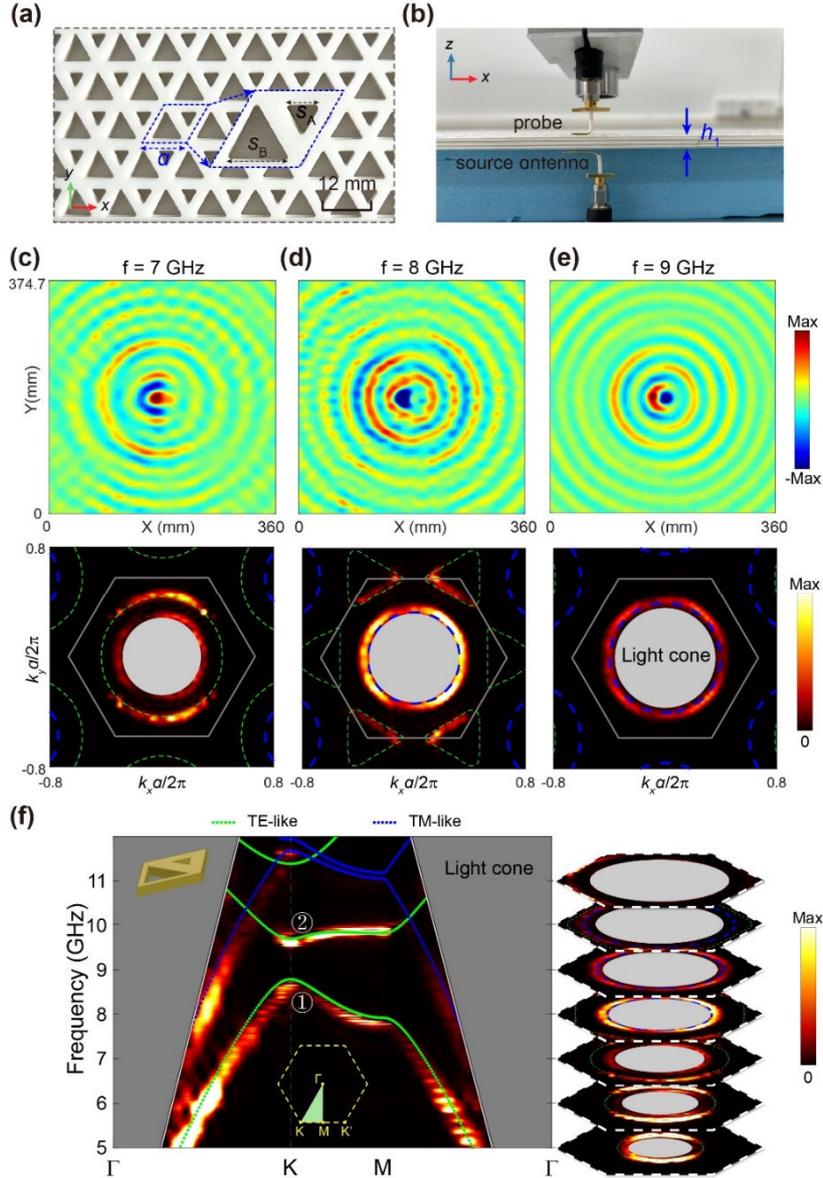

**FIG. 2. Bulk bands of the monolayer VPC.** (a) Photograph of the monolayer VPC slab whose unit cell consists of two triangular air holes with different sizes. (b) Experimental setup to excite and probe the electric fields. A bent source antenna is placed beneath the slab while another probe antenna scans the upper surface of the slab. (c-e) Measured $E_x$ fields and the corresponding eigenfrequency contours at three discrete frequencies (i.e. 7 GHz, 8 GHz, and 9 GHz). Green and blue dashed lines represent the simulated eigenfrequency contours of the TE-like and TM-like bulk modes, respectively. The light cone is shaded in grey. Notably, the contours of TE-like bulk modes show a closed circle around the Γ point, form triangle-shaped around the K and K' points, and disappear with the increasing of frequency. (f) The bulk band structure along high symmetry $k$-points in the first Brillouin zone. Right panel shows the eigenfrequency contours at different frequencies. Clearly, a band gap is found between the first and second TE-like bulk bands.

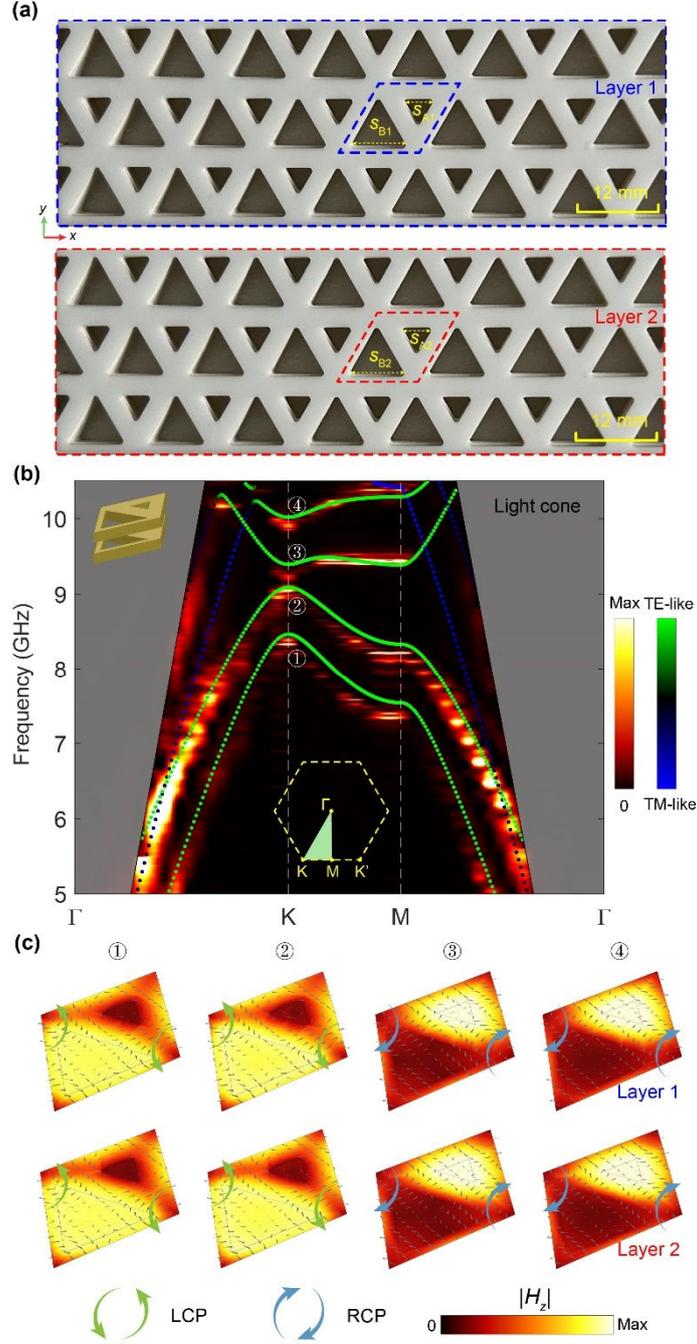

**FIG. 3. Bulk bands of the mirror-symmetric bilayer VPC.** (a) Photographs of the upper Layer 1 and the lower Layer 2 of the MS bilayer VPC. The lateral dimensions of the air holes are $s_{A1} = s_{A2} = 3.6$ mm, $s_{B1} = s_{B2} = 8.4$ mm. (b) Measured bulk bands along high symmetry $k$-points. The simulated results are represented by green dots (TE-like modes) and blue dots (TM-like modes), respectively. Around the K point, four TE-like bulk bands are separated due to the mode coupling between eigen modes with two layers. (c) Simulated $|H_z|$ fields at the central plane of Layer 1 and Layer 2 of four bulk modes at the K point. The color shows the $|H_z|$ fields while gray arrows denote the energy flux. The green and blue arrows indicate the rotation direction of the energy flux, indicate that mode ① and mode ② belong to LCP orbital states, while mode ③ and mode ④ belong to RCP orbital states

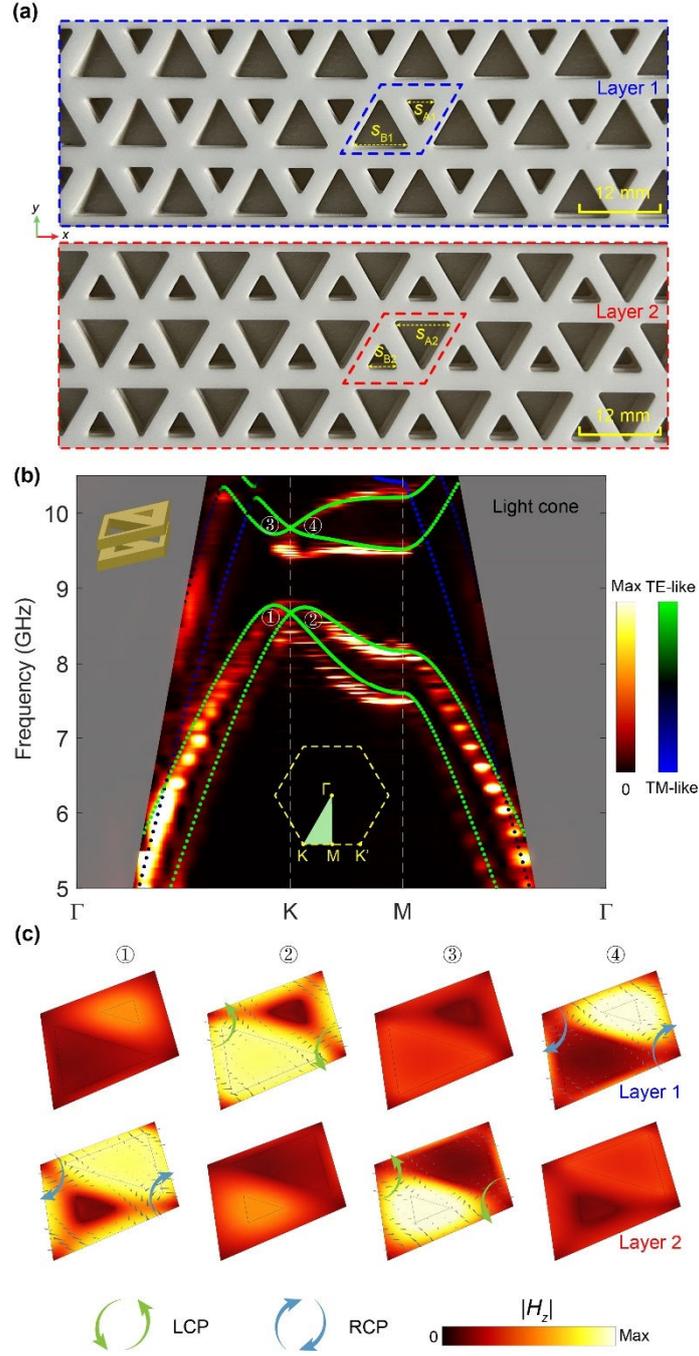

**FIG. 4. Bulk bands of the inversion symmetric bilayer VPC. (a)** Photographs of the upper Layer 1 and the lower Layer 2 of the IS bilayer VPC. The lateral dimensions of the air holes are $s_{A1} = s_{B2} = 3.6$ mm, $s_{B1} = s_{A2} = 8.4$ mm. **(b)** Measured bulk bands along high symmetry k-points. The simulated results are represented by green dots (TE-like modes) and blue dots (TM-like modes), respectively. Under the protection of inversion symmetry, the first and second (third and fourth) bulk band are degenerate at the K point. **(c)** $|H_z|$ fields at the central plane of Layer 1 and Layer 2 of four bulk modes at the K point. The eigen fields are localized in different layers, indicating the photonic pseudospin orbital coupling.

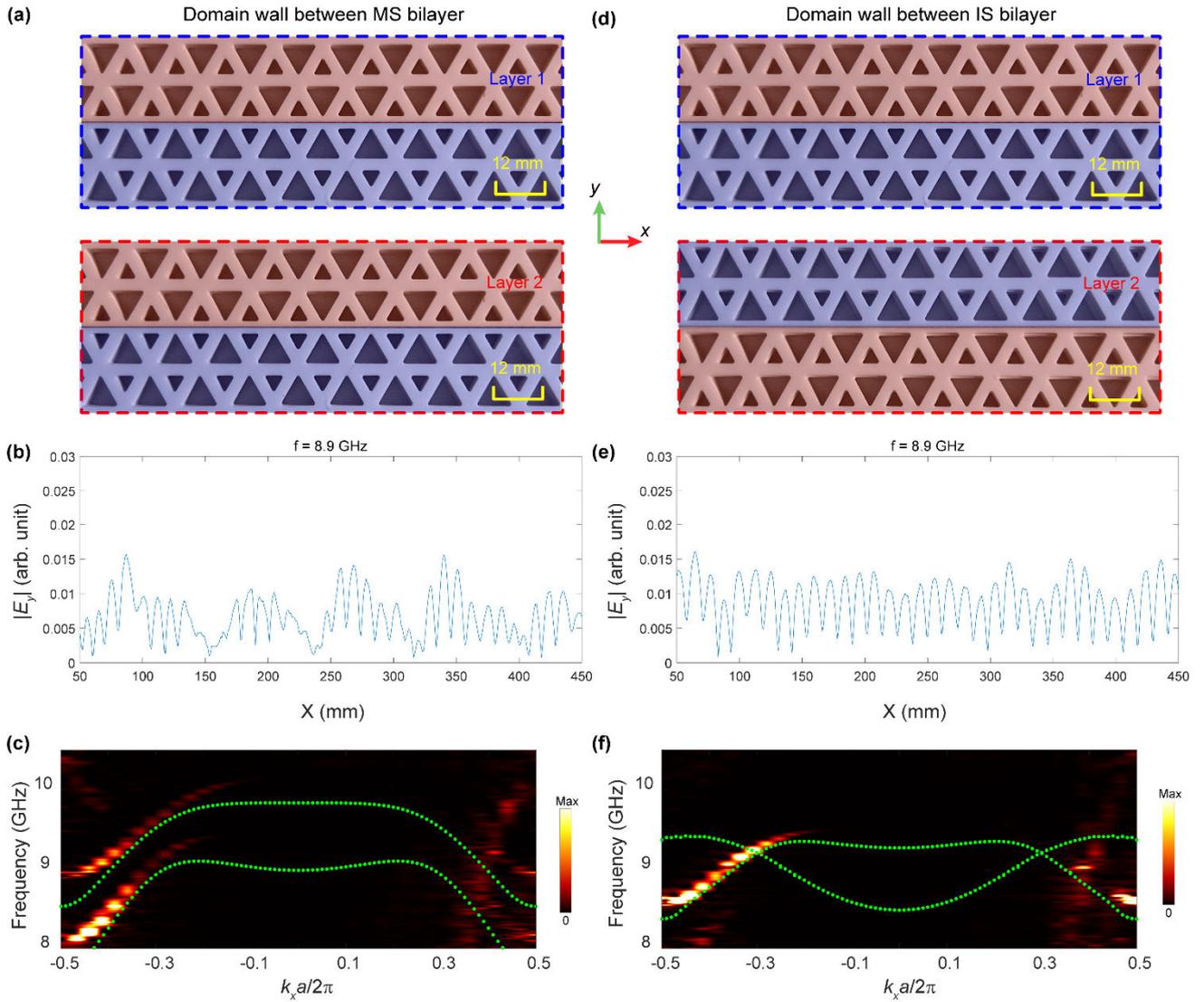

**FIG. 5. Experimental probing of layer-mixed and layer-polarized edge modes. (a)** Magnified view of the domain wall formed by the MS bilayer VPCs. The Layer 1 and Layer 2 slabs have the same boundary configuration. **(b)** Measured $|E_y|$ fields along the domain wall between MS bilayer VPCs when the source antenna is placed at the left side of the domain wall. **(c)** Measured edge dispersion and the simulated results are highlighted by green dots. Two edge dispersions with the same group velocity are found. **(d)** Magnified view of the domain wall formed by the IS bilayer VPCs. The Layer 1 and Layer 2 slabs have different boundary configuration. **(e, f)** Measured $|E_y|$ fields along the domain wall between MS bilayer VPCs and the corresponding edge dispersion. Only one edge dispersion is measured as the layer-polarized edge mode has its field localized within a single layer.